\documentclass[aps,prd,twocolumn,groupedaddress,nofootinbib]{revtex4}
\pdfoutput=1
\usepackage{graphics}
\usepackage{braket}
\usepackage{amsmath}
\usepackage[utf8]{inputenc}
\usepackage{graphicx}
\usepackage{cleveref}

\newcommand{\defeq}{\stackrel{\text{def}}{=}}
\begin{document}

\title{Localisation transition in the driven Aubry-André model}
\author{D. Romito}
\author{C. Lobo}
\affiliation{University of Southampton}
\author{A. Recati}
\affiliation{Università degli Studi di Trento}

%\date{\today}
% Remove option referee for final version
%
% Remove any % below to load the required packages
%\usepackage{latexsym}

\begin{abstract}
A recent experiment by P. Bordia et al. \cite{bloch} has demonstrated that periodically modulating the potential of a localised many-body quantum system described by the Aubry-André Hamiltonian with on-site interactions can lead to a many-body localisation-delocalisation transition, provided the modulation amplitude is big enough.
Here, we consider the noninteracting counterpart of that model in order to explore its phase diagram as a function of the strength of the disordered potential, the driving frequency and its amplitude.
We will first of all mimic the experimental procedure of \cite{bloch} and use the even-odd sites imbalance as a parameter in order to discern between different phases. 
Then we compute the Floquet eigenstates and relate the localisation-delocalisation transition to their Inverse Participation Ratio. 
Both these approaches show that the delocalisation transition occurs for frequencies that are low compared to the bandwidth of the time independent model. Moreover, in agreement with \cite{bloch} there is an amplitude threshold below which no delocalisation transition occurs.
We estimate both the critical values for the frequency and the amplitude.
\end{abstract}

\maketitle
\section{Introduction}
\label{intro}
The study of periodically driven quantum systems has gained interest in the last years.
The most important tool in this context is the \textit{Floquet theorem}.
One of its most interesting consequences is that it allows to write the evolution over multiples of the driving period in terms of a time independent effective Hamiltonian \cite{floshi}, \cite{flosam}. 
The existence of such an effective Hamiltonian can open the way to the so called \textit{Floquet engineering}, that is, the possibility to realise non trivial time independent models by periodically modulating a quantum system with a suitable protocol. This concept has been employed very successfully in various experiments with ultracold atoms in driven optical lattices \cite{eckrev}. This includes dynamic localisation (\cite{dundyn}, \cite{holdyn}, \cite{gridyn}, \cite{ligdyn}), “photon”-assisted tunneling (\cite{zakphot}, \cite{eckphot}), control of the bosonic superfluid-to-Mott-insulator transition (\cite{ecksup}, \cite{zensup}) and the realisation of artificial magnetic fields (\cite{haldane}, \cite{aidelsburgerstrong}, \cite{tungauge}, \cite{Daliba}).

A recent seminal experiment \cite{bloch} has focused instead on the case of a driven many-body localised system, described by the Aubry-André model \cite{auband}, showing that a properly tuned periodic driving can lead the system across the localisation transition.
The interplay between a many-body localised quantum system and a periodic driving has drawn a lot of attention and recent theoretical works have put forward the possibility that this combination can generate symmetry protected topological phases which have no equilibrium analogues \cite{topological1} \cite{topological2} \cite{topological3} \cite{topological4} \cite{topological5}.
Motivated by the particular case of the experiment reported in \cite{bloch} we consider its single particle counterpart, i.e. the periodically driven Aubry-André model, in order to understand which kind of effects can be disentangled from a strict many-body description.
The same model was recently studied in \cite{Sinha}, with which our results show good agreement, particularly in estimating the critical amplitude.

\section{The model}
\label{sec:model}
We consider the Aubry-André Hamiltonian $H_0$, with periodically modulated potential $V(t)$:
\begin{equation}
 H(t)=H_0+V(t)
\end{equation}
where

\begin{equation}
\label{H_0}
H_{0}=J\sum_i^{N} \left(\ket{i}\bra{i+1}+\ket{i+1}\bra{i}\right)+\lambda \sum_i\cos(2\pi \beta i +\phi)\ket{i}\bra{i}
\end{equation}

\begin{equation}
V(t)=A\cos (\omega t) \sum_i \cos(2\pi \beta i+\phi)\ket{i}\bra{i}
\end{equation}
Here $\beta$ is an irrational number, $\ket{i}$ is the Wannier state on site $i$ and $\lambda$ is the disorder strength. We choose periodic boundary conditions for the lattice.
The term $V(t)$ satisfies $V(t+T)=V(t)$, where $T=2 \pi/\omega$.

The qualitative features of the undriven system in the many-body and the noninteracting cases are quite similar \cite{praund}.
In the noninteracting case the Hamiltonian $H_0$ is that of a particle moving in a lattice with two lenghts $L_1$ and $L_2$ which are incommensurate, with $\beta=L_1/L_2$.
This produces a pseudorandom potential and the different realisations of this randomness are obtained changing the value of the phase $\phi$.
Such a Hamiltonian is the simplest realisation of a quasiperiodic crystal, or quasicrystal (see \cite{revQC} for a review on the topic).
%In what follows we will indicate the disorder strength, $\lambda$, and the amplitude of the modulation, $A$, in units of $J$ and times in units of $1/J$.
When $\beta$ is an irrational number the time independent model is proven to undergo a metal to insulator transition: above $\lambda=\lambda_c=2J$ the eigenstates of $H_0$ are localised, whereas below they are delocalised (\cite{auband}, \cite{Azbel}, \cite{Aulbach})

In the many-body case the transition is controlled by the parameters $J$, $\lambda$ and $U$, where $U$ is the intensity of the repulsive on-site interaction. 
There is a critical disorder strength which depends on $J$ and $U$ above which the system becomes localised \cite{timeindependent}. More precisely, until $U \approx 2\lambda $ the interaction decreases the degree of localisation, while for large $|U|$, increasing $U$ helps to make the system more localised. 
This is understood as a consequence of the formation of repulsive stable bound atom pairs in optical lattices described by a Hubbard Hamiltonian (\cite{Mattis}).For the first realization of this effect with cold atoms see \cite{Hubbard} These pairs have a reduced effective tunneling rate of $J_{eff} \approx J^2/|U|$ which thus increases the degree of localisation .
Both above and below $2\lambda$ for each value of $U$ there is a definite value of $\lambda$ for which the transition occurs.
It is thus interesting to see whether these analogies are retained in presence of the time periodic modulation.

\section{Setup}
\label{sec:setup}
\subsection{Imbalance}
As a first step to explore the phase diagram of the model we mimic as closely as possible the procedure described in \cite{bloch} but in a single particle context.
The initial state there is chosen as a density-wave pattern in which fermions occupy the even sites of the lattice. 
The parameter which discerns between a localised and a non-localised phase is the asymptotic imbalance:
\begin{equation}
\label{imbalance}
I=\lim_{t \rightarrow \infty} \frac{1}{t} \int_0^t dt' \dfrac{N_e(t')-N_o(t')}{N_e(t')+N_o(t')}
\end{equation}
where $N_e$ and $N_o$ are the number of particles in the even and odd sites respectively.
A persistent imbalance indicates a localised phase, while it obviously drops to 0 in absence of localisation, indicating that the system is ergodic as it does not retain the memory of its initial conditions.

To properly imitate the experiment we consider different realisations of the system each initially localised on a single even site and let them evolve separately under the Hamiltonian $H(t)$. The initial state in the Wannier states basis for each realisation $m=1,...,N/2$ reads:
\begin{equation}
\psi^{(2m)}(i,t=0)=\braket{i|\psi^{(2m)}(t=0)}=
   \delta_{i,2m} %\hspace{1cm} \text{for } m=1,...,N/2
\end{equation}
After a long evolution time we sum the modulus squared amplitudes of all the realisations to obtain the final density, namely:
\begin{equation}
n(i,t)=\sum_{m=1,...N/2} |\psi^{(2m)}(i,t)|^2
\end{equation}
The above definition is justified by the fact that the one-body density of a noninteracting many-body system is the sum of the densities of the occupied single particle states \cite{manybodybook}.
The analogues to the occupation numbers $N_e$ and $N_o$ are then calculated by simply using this density function as a weight in the following sum:
\begin{equation}
N_e(t)=\sum_{i=1}^{N/2} n(2i,t)% \delta_{i,2m}
\end{equation}
and similarly
\begin{equation}
N_o(t)=\sum_{i=0}^{N/2-1} n(2i+1,t)% \delta_{i,2m+1}
\end{equation}
With these definitions we can simply calculate the imbalance as defined in equation \ref{imbalance}.

\begin{figure}

\includegraphics[width=8.8cm]{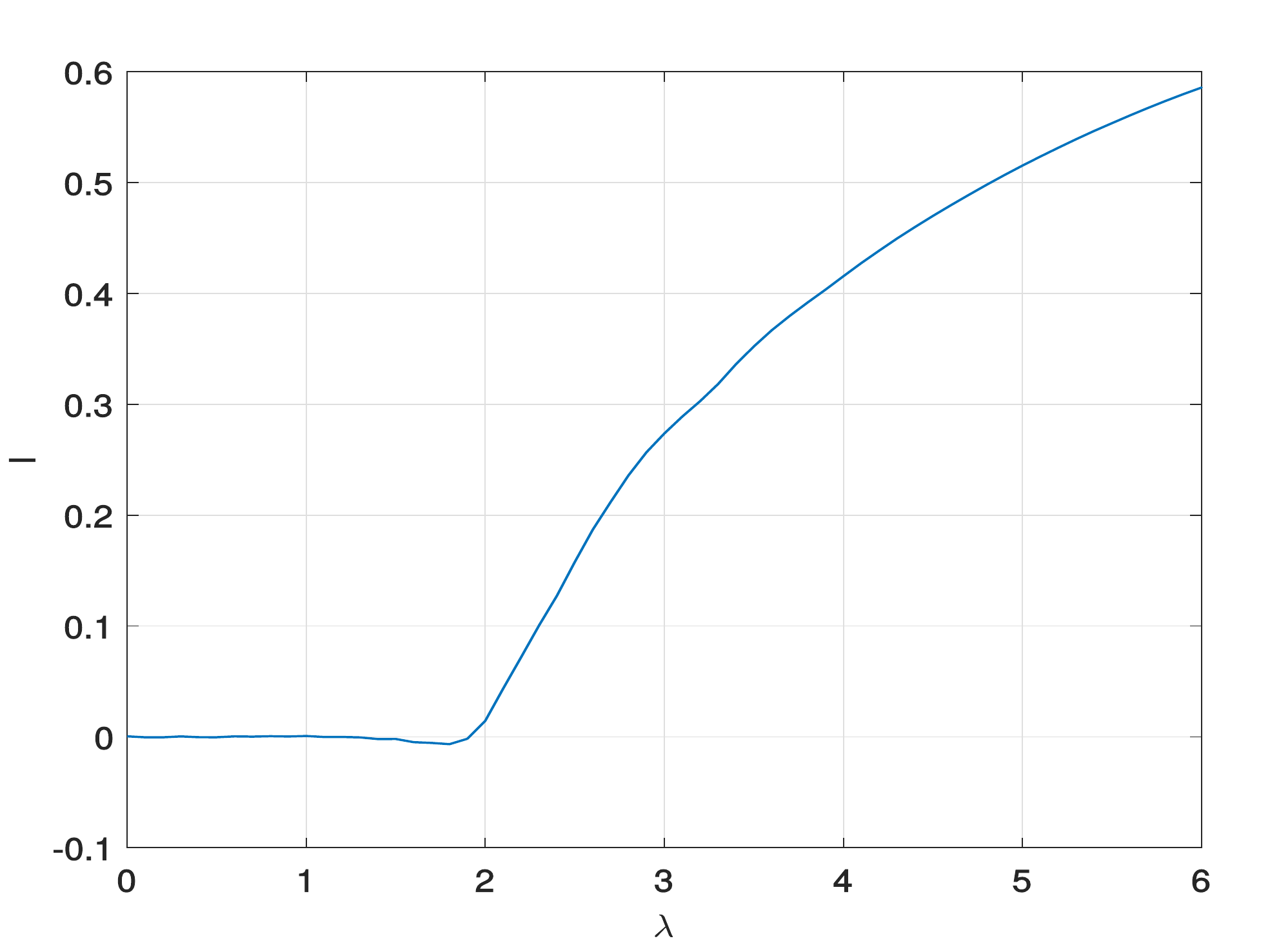}
\caption{Imbalance in the time independent case as a function of the disorder strength $\lambda$, for lattice size $N=50$, with periodic boundary conditions.
We see a clear critical value at $\lambda_c=2J$ indicating the localisation transition. The time of integration is $\tau=1000 \hbar/J$, at which the imbalance has reached its asymptotic value}
\label{timeindependent}
%\floatfoot{Asymptotic imbalance for the time independent case as a function of the disorder strength $\lambda$ (in units of $J$).
%The asymptotic imbalance is calculated as the time averaged imbalance in the second half of the evolution, and it's averaged over 20 different realisations of disorder, for a lattice made of $N=50$ sites}\label{timeindependent}
\end{figure}

Before moving to the results for the driven lattice we show how the Imbalance behaves around the phase transition for the time independent model.
Figure \ref{timeindependent} was obtained considering a lattice made of $N=50$ sites and calculating the asymptotic Imbalance for different values of the disorder strength $\lambda$.
It shows how the transition is marked by a nonzero value of the imbalance as a function of $\lambda$ (all energies are in units of $J$), at the critical value $\lambda_c=2J$.

The imbalance is thus able to signal in a clear way the transition from the localised to the delocalised phase.

\subsection{Inverse Participation Ratio}
In this subsection we link the localisation properties of the model to the localisation of its Floquet modes.
The time periodicity of the full Hamiltonian $H(t)$ allows us to make use of the Floquet theorem, which states that we can write the time evolution of an arbitrary initial state as \cite{floshi}, \cite{flosam}:
\begin{equation}
\label{eq:floquet}
\ket{\psi(t)}=\sum_n c_n e^{-i\epsilon_n(t)/\hbar}\ket{u_n(t)}
\end{equation}
with $c_n=\braket{u_n(0)|\psi(0)}$ and $\epsilon_n$ are the so-called quasienergies. We emphasize the fact that these coefficients do not depend on time.
The states $\ket{u_n(t)}$ are called Floquet modes.
They have the same time periodicity of the original Hamiltonian and can be found as eigenstates of the propagator over one period, which reads:
\begin{equation}
U(T,0):=\mathcal{T} \exp{\bigl(-\cfrac{i}{\hbar} \int_{0}^T H(\tau) d\tau \bigr)}
\end{equation}
where $\mathcal{T}$ denotes the time ordering operator.

We expand the Floquet modes at $t=0$ in the Wannier state basis yielding:
\begin{equation}
\ket{u_n(0)}=\sum_{i} b_i^{(n)} \ket{i}.
\end{equation}
We define the averaged Inverse Participation Ratio (IPR) as the average of the IPRs of all the Floquet eigenmodes on the Wannier states, namely:
\begin{equation}
\text{IPR}=\frac{1}{N} \sum_{i,n} |{b_i^{(n)}}|^4
\end{equation}
where $N$ represents the number of Floquet modes which coincides with the number of sites of the lattice.

If each one of the Floquet eigenmodes is localised on a single Wannier state then for any $n$ there exists an $i$ such that $|{b_i^{(n)}}|\approx 1$ and the sum approaches $1$. If instead the eigenmodes are distributed among many Wannier states then $|{b_i^{(n)}}| \approx 1/\sqrt{N}$ for all $n$, $i$ and the averaged IPR goes to $0$ as $1/N$.

Thanks to the form of equation (\ref{eq:floquet}) we can expect a localised dynamics when very few Floquet modes participate in the time evolution of an initial Wannier state.
This would resemble in the context of time-periodic systems the phenomenon of Anderson localisation where the localisation of the eigenstates of the Hamiltonian implies non-ergodic dynamics \cite{andloc}.
This is not however the only mechanism that is in play, as localisation can occur as a consequence of the degeneracy of energy levels e.g. when the renormalized hopping parameter $J$ becomes very small due to the driving. This particular mechanism is often referred to as dynamic localisation or band collapse (\cite{DreseHolthaus}, \cite{eckdyn}). In \cite{DreseHolthaus}, in particular, the authors propose to observe a dynamic localisation effect in a realization of the Aubry-André Hamiltonian by tuning the amplitude and frequency to a value for which the renormalized hopping would vanish. The periodic modulation that we are considering here is however different from theirs and doesn't allow the same tunability of the renormalized hopping. 
Confronting the Inverse Participation Ratio with the Imbalance allows us to discern which effects are due to collapse of the bandwidth of quasienergies and which ones are due to an Anderson localisation transition.

\section{Results}
In what follows we will indicate the disorder strength, $\lambda$, and the amplitude of the modulation, $A$, in units of $J$ and times in units of $1/J$.
The calculations below were done considering a lattice made of $N=50$ sites, averaging over $20$ different realisations of the disorder, which are obtained by varying the value of the phase $\phi$ in equation (\ref{H_0}).
In choosing $\beta$ we decided to follow as close as possible the choice of the experiment in reference [1], so we chose $\beta=532/738.2$.
The simulations were made using the standard Matlab toolbox, solving the time evolution with the ode45 function in order to compute the imbalance and exactly diagonalizing the propagator over one period to find the Floquet modes.

\begin{figure}
\includegraphics[width=\linewidth]{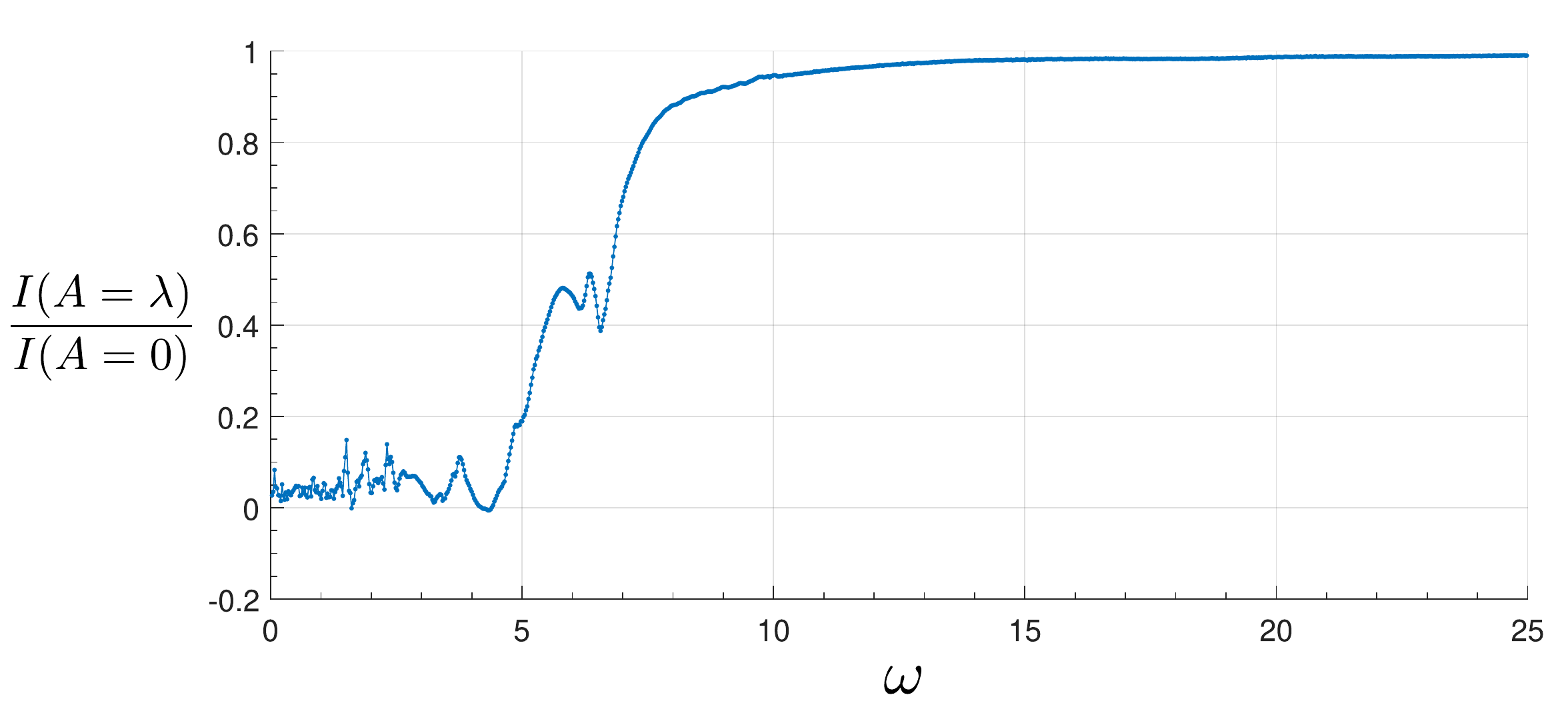}
\caption{Imbalance as a function of frequency for $A=\lambda$, normalized to its value in the absence of driving. While for low frequencies the imbalance is vanishing, it approaches the value for $A=0$ for high frequencies. This plot can be compared with the ones present in \cite{bloch} and confirms the fact that the single particle picture is able to capture the qualitative features of the many-body experiment.}
\label{Ivsfreq}
\end{figure}

As a first step to outline the behaviour of this model we calculated the imbalance for strong driving, i.e. $A=\lambda$, for a broad range of frequencies, keeping the disorder strength at a fixed value above the critical one $\lambda=3=A$.
The results are shown in figure \ref{Ivsfreq}, which highlights a delocalised regime for low frequencies while for high frequencies the imbalance approaches that of the model in absence of driving.
The similarity in the main features between this figure and the ones in \cite{bloch} is striking. In particular the dip appearing after the imbalance has started to rise, around $\hbar \omega= 2\lambda$, is present also in the many-body experiment.

\begin{figure}
\includegraphics[width=\linewidth]{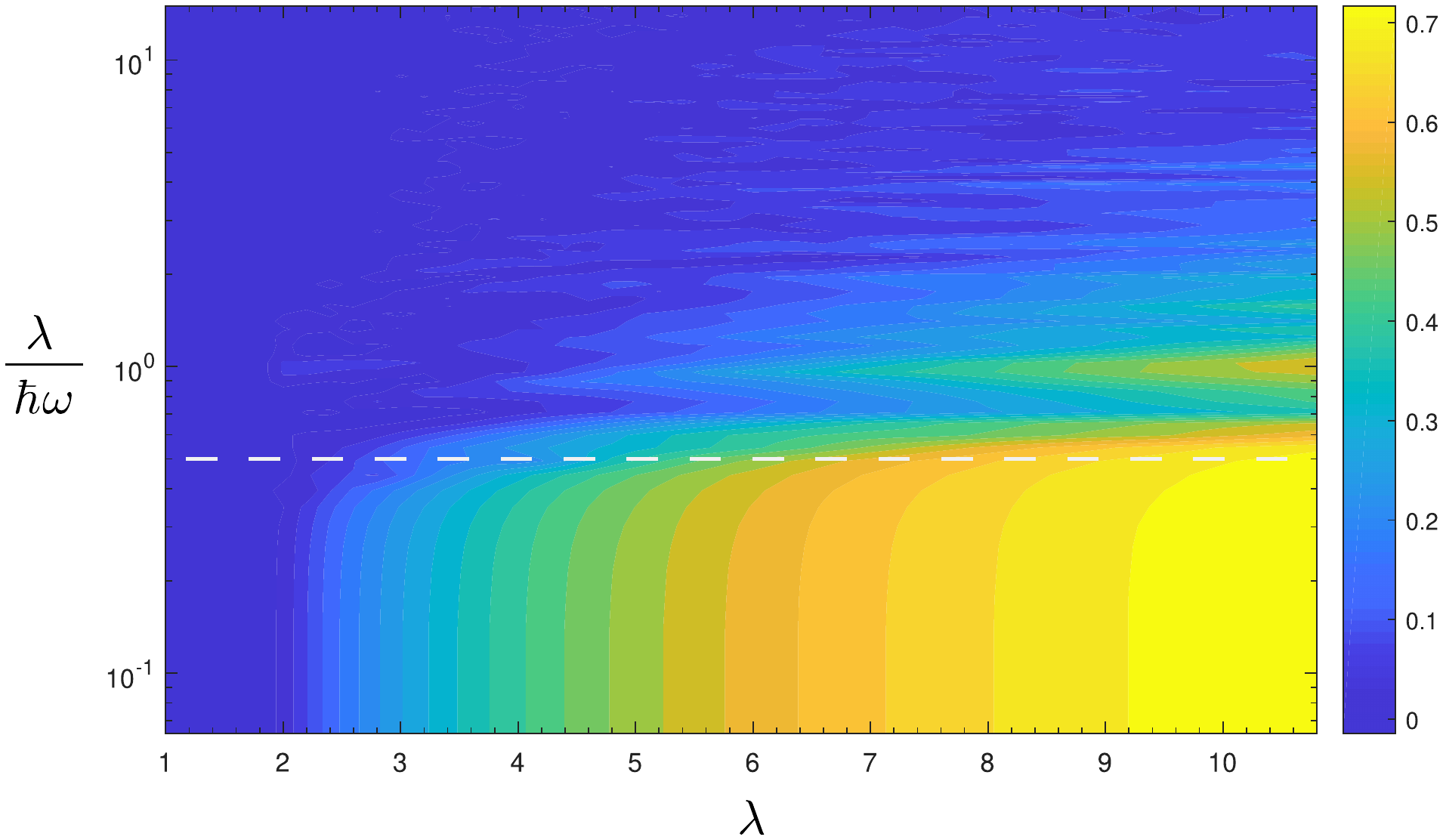}
\caption{Imbalance as a function of frequency and disorder strength. On the vertical axis $\frac{\lambda}{\hbar \omega}$ is displayed in logarithmic scale to allow comparison with the experiment in \cite{bloch}. The dashed line is for $\hbar \omega = 2\lambda$, which is the approximate critical value for the frequency.}
\label{frequencyImbalancelog}
\end{figure}

\begin{figure}
\includegraphics[width=\linewidth]{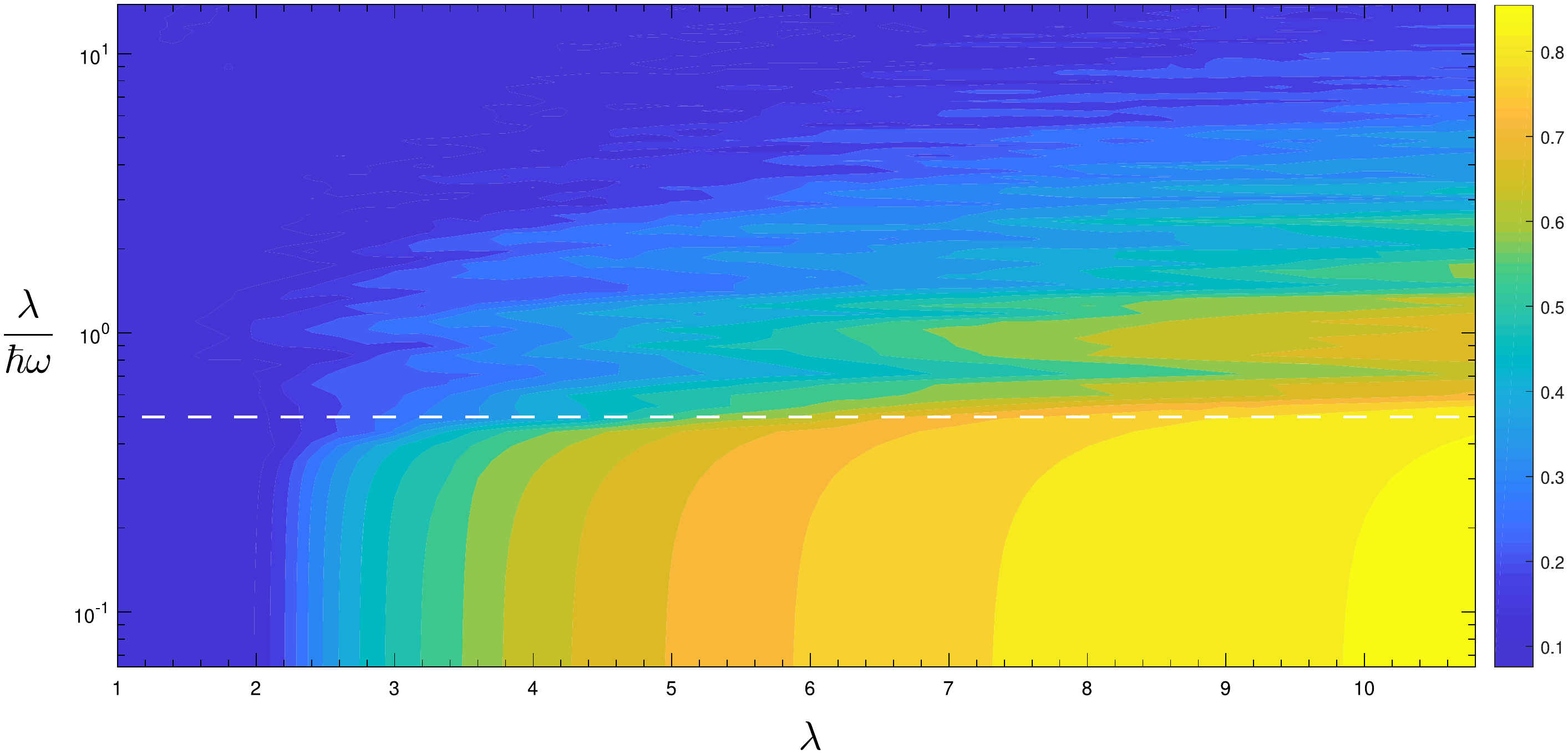}
\caption{Inverse Participation Ratio as a function of frequency and disorder strength. On the vertical axis $\frac{\lambda}{\hbar \omega}$ is displayed in logarithmic scale to allow comparison with the experiment in \cite{bloch}. The dashed line is for $\hbar \omega = 2\lambda$, which is the approximate critical value for the frequency.}
\label{frequencyIPRlog}
\end{figure}

\begin{figure}
\includegraphics[width=\linewidth]{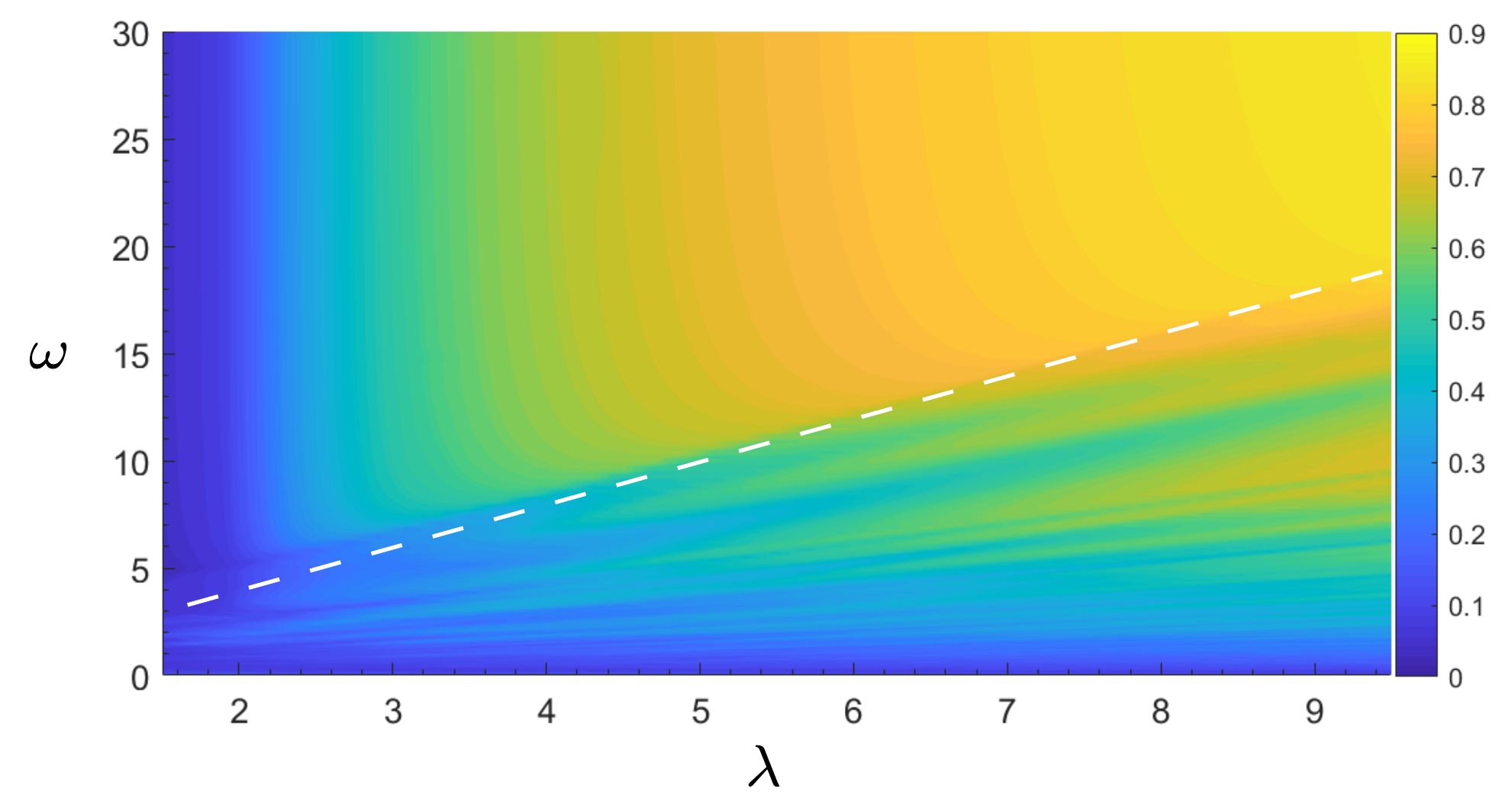}
\caption{Inverse Participation Ratio as a function of frequency and disorder strength, for $A=\lambda$. The white dashed line is for $\hbar \omega = 2\lambda$, dividing the localised phase (yellow) from the delocalised one (blue).}
\label{frequencyIPR}
\end{figure}

To explain the origin of the above mentioned dip in the value of the imbalance we first have to consider the response of the model to various values of frequency and disorder.
To this intent we computed the time averaged imbalance for different values of the disorder strength $\lambda$ and the angular frequency $\omega$, setting the amplitude of the modulation in the strong driving regime i.e. $A=\lambda$.
The evolution time is chosen to be $100$ times the period of the modulation.
In \ref{frequencyImbalancelog} and \ref{frequencyIPRlog} the vertical axis shows $\frac{\lambda}{\hbar \omega}$ to allow comparison with the experiment in \cite{bloch}.
In \ref{frequencyIPR} the Inverse Participation Ratio is displayed as a function of $\lambda$ and $\hbar \omega$ to more clearly show the relation between the frequency response and the spectrum.

%\begin{figure}
%\includegraphics[width=\linewidth]{frequency(I)_new.pdf}
%\caption{Imbalance as a function of frequency and disorder strength, for $A=\lambda$. The white dashed line is for $\hbar \omega = 2\lambda$, dividing the localised phase (yellow) from the delocalised one (blue).}
%\label{frequencyImbalance}
%\end{figure}

Figure \ref{frequencyImbalancelog} confirms that for very low frequencies the system is brought to a delocalised phase (marked by a vanishing imbalance), while for high frequency the driving is not able to bring the system to delocalisation anymore. Moreover it displays distinct analogy with the corresponding phase diagram in \cite{bloch}.

As anticipated in the previous section we computed the Inverse Participation Ratio for various values of $\omega$ and $\lambda$. Figures \ref{frequencyIPR} and \ref{frequencyIPRlog} distinctly show the separation between the two phases. This also shows that the localisation properties of the Floquet eigenmodes at initial time allow us to discern in a broad sense the different localisation properties of the system.
This happens despite the fact that the initial Floquet modes carry no information on the structure of the quasienergy spectrum which can contain accidental crossings of energy levels, causing the system to be partially localised.
The relatively small size of the system implies that the Inverse Participation Ratio will display finite size effects in the delocalised phase, where it vanishes as $1/N$. We run a simulation which computes the IPR as a function of frequency for different system sizes, going from $N=50$ to $N=500$. For each system size the IPR goes to $0$ with the correct scaling with respect to $N$, while in the localised phase its behaviour is largely unchanged.

\begin{figure}
\includegraphics[width=\linewidth]{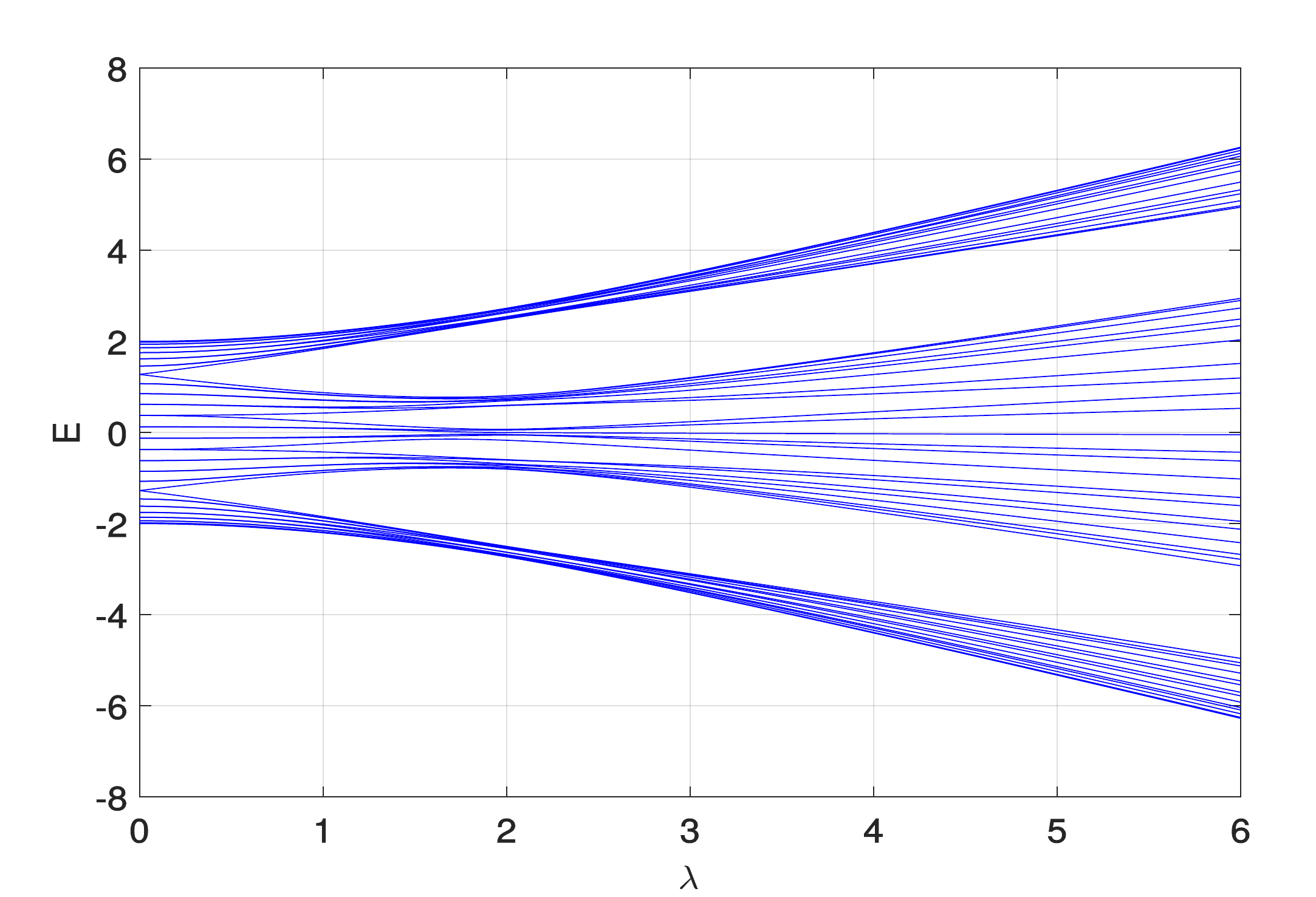}
\caption{Spectrum of the Aubry-André model as a function of $\lambda$ for $N=50$ lattice sites.}
\label{fig:spectrum}
\end{figure}

There's a transition line (white dashed line in \ref{frequencyImbalancelog}, \ref{frequencyIPRlog} and \ref{frequencyIPR}) above which the system remains localised. This line appears for $\hbar \omega_c=2\lambda$ which can be understood from the spectral properties of the Hamiltonian $H_0$ of equation \ref{H_0}.
To better illustrate this, in figure \ref{fig:spectrum} we show the spectrum of the Aubry-André model as a function of the disorder strength for $N=50$ lattice sites.
The bandwidth of the Aubry-André Model is $\approx 2\lambda$ for any disorder strength above the transition point $\lambda_c=2J$.
Thus the transition line in the time periodic case appears when the quanta of energy that the driving pumps into the system are too big for the system to absorb.
Above the transition line the system's behaviour becomes that of the time independent model.
This is because the period of the driving $T=\frac{2 \pi}{\omega}$ is now smaller than the fastest time scale present in the Aubry-André Hamiltonian, making the system unable to respond to the driving.

Below the transition line there are other smaller revivals of the localised phase. We attribute this intricate structure again to the spectrum of the Aubry-André model which is divided into smaller subbands divided by spectral gaps. In the intermediate range of frequencies where $\hbar \omega$ is comparable to the energy gaps present in the spectrum, the presence of a localised phase has a non monotonic dependence on the frequency of the modulation.
The study of the relevance of the single particle spectrum and the density of spectral lines for the frequency response of the model was studied in the thesis that lead to this paper \cite{thesis}.

All these results confirm the qualitative picture outlined in \cite{bloch} and are well understood in terms of the single particle spectrum. This suggests that in this context the time averaged imbalance, while providing a precise characterization of the phase diagram of the model, doesn't seem able to disentangle the role of the interactions.

For very low frequencies the system is brought to delocalisation. We can define the following parameter:
\begin{equation}
\tilde{\lambda}(t) \defeq \lambda+A\cos(\omega t)
\end{equation}
If the frequency is very low the global parameter $\tilde{\lambda}(t)$ is changed adiabatically and sweeps through the transition point $\lambda_c=2J$, bringing the system to delocalisation.

This intuitive picture helps us understand the role of the amplitude of the driving $A$: even for arbitrarily low frequencies the system does not delocalise if the amplitude is not big enough to make $\tilde{\lambda}(t)$ sweep through the critical point $\lambda_c=2J$.
Following this reasoning we define the critical value for $A$ to be such that $\min_t \left\{ \tilde{\lambda}(t) \right\} =\lambda_c=2J$, namely:
\begin{equation}
\label{eq:Acrit}
A_c=\lambda-\lambda_c=\lambda-2J
\end{equation}

\begin{figure}
\includegraphics[width=\linewidth]{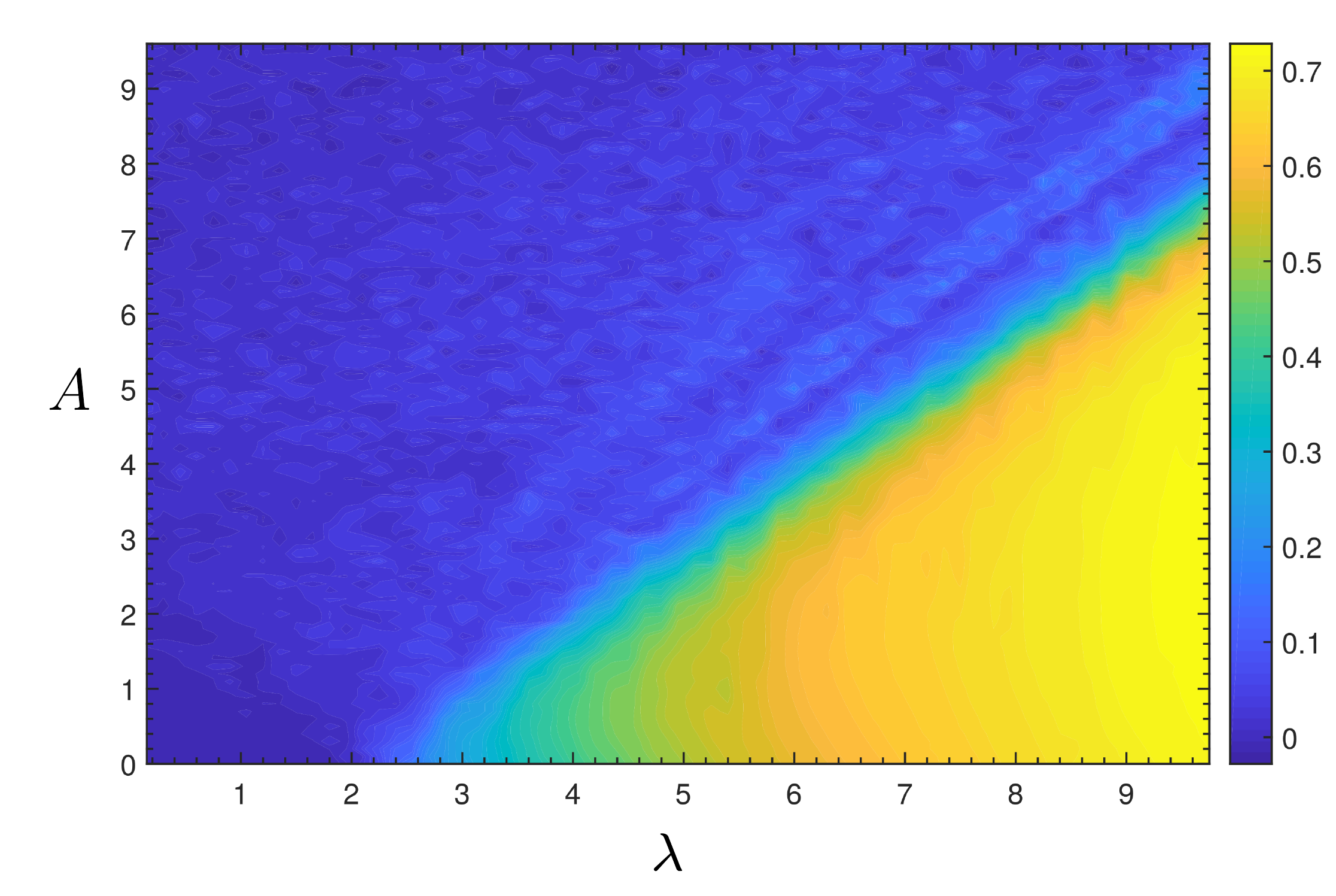}
\caption{Imbalance as a function of amplitude and disorder strength, for $\nu=0.005(1/J)$. There's a clear line for $A=\lambda-2$ separating the localised phase (yellow) to the delocalised one (blue).}
\label{fig:Acrit(I)}
\end{figure}

\begin{figure}
\includegraphics[width=\linewidth]{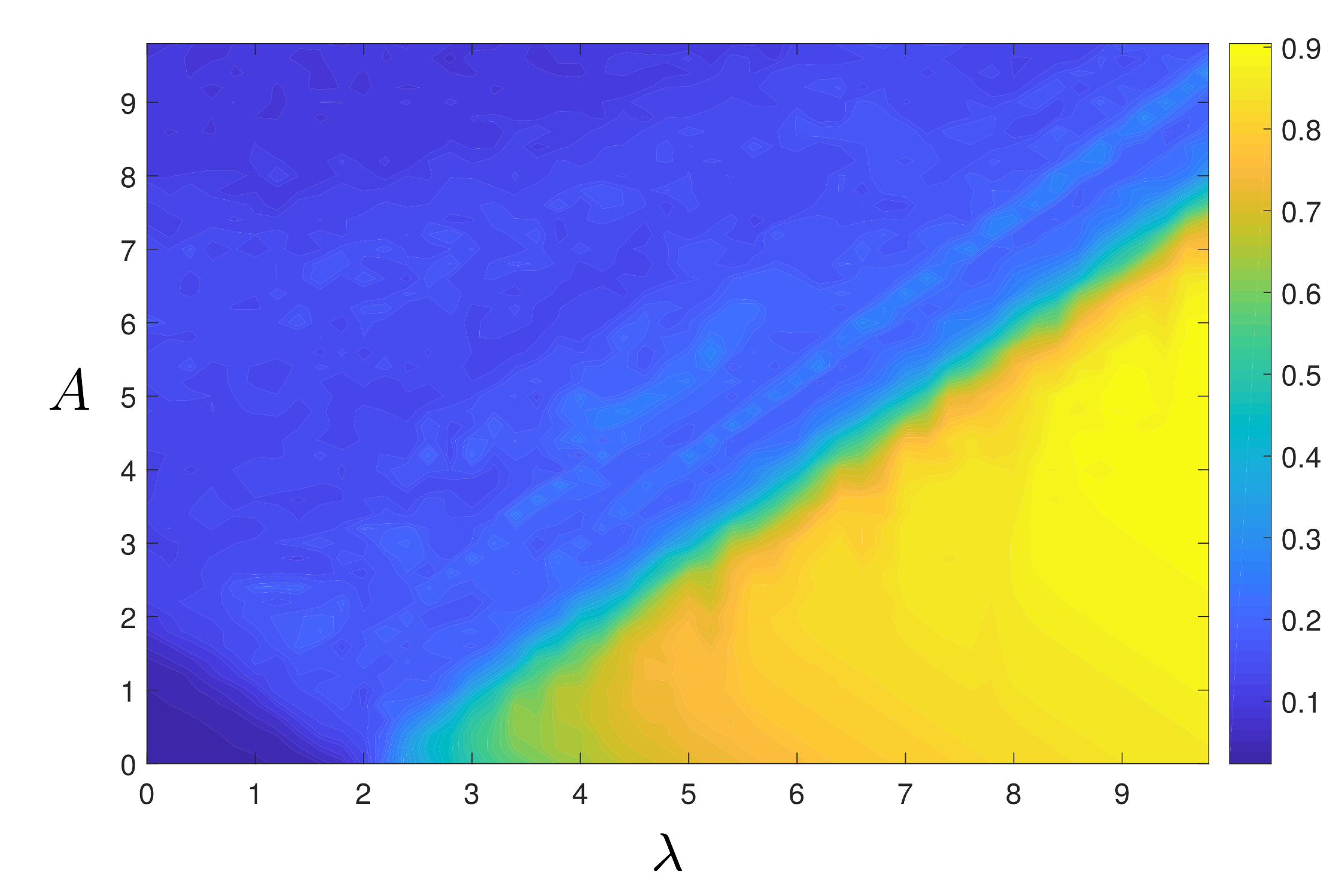}
\caption{Inverse Participation Ratio of the Floquet eigenmodes as a function of amplitude and disorder strength, for $\nu=0.005(1/J)$. The result is consistent with the one in fig. \ref{fig:Acrit(I)}, confirming that the localisation properties of the model are due to the localisation of the Floquet eigenmodes}
\label{fig:Acrit(IPR)}
\end{figure}

This picture is clearly confirmed by the contourplots of the imbalance and the Inverse Participation Ratio as functions of the disorder strength and the amplitude, which are shown on figure \ref{fig:Acrit(I)} and \ref{fig:Acrit(IPR)}. For these plots we considered a frequency $\nu=\omega/2\pi=0.005 (1/J)$.
The same value for the critical amplitude was found in \cite{Sinha}.

We stress that the very existence of a critical value of $A$ as determined here is valid only in the case of a modulation of the form considered in this work, which corresponds to a modulation of the disorder strength.
It is often stated in the literature (see \cite{abanin}, \cite{regimes}, \cite{rehn}) that a modulation of arbitrarily small amplitude will always delocalise a many-body localised quantum system provided that the driving frequency is small enough.
This statement however is not in contrast with our result as it refers to a driving of the form $A\cos(\omega t) \sum_i i \ket{i} \bra{i}$.

Last, we  want to briefly comment on the role of the interaction. In presence of the interaction $U$, we expect to retain most of the qualitative results displayed, as comparison with the many-body experiment seems to suggest. In particular, regarding the frequency response of the system we expect the description in terms of the spectrum to be still relevant, with the critical value $\omega_c$ to be shifted to be equal to the badwidth of the interacting model. According to \cite{mastro} the many-body interaction will change the size of the infinite number of spectral gaps of the noninteracting spectrum, without closing any of them. This makes the interacting model very similar to the noninteracting one, except for the intermediate frequency regime where the effect of the interaction in coupling the energy levels is crucial, possibly explaining the less sharp peaks displayed in the experiment in \cite{bloch}.

\section{Conclusions}
Our work shows that the driven noninteracting Aubry-André model qualitatively reproduces many of the localisation phenomena which are found in the experiment \cite{bloch} such as the presence of a delocalised phase for low frequency, the persistence of localisation for a high frequency driving, and the existence of a critical value of driving amplitude for the onset of the localisation transition.
We were able to determine the critical values for the frequency and the amplitude of the driving, and provide a theoretical explanation for their existence.
Moreover we related the phases of the model to the localisation of its Floquet eigenmodes.
Future theoretical studies should focus on the role of interactions and the new qualitative aspects they bring.


%\resizebox{0.75\textwidth}{!}{%
  %\includegraphics{leer.eps}
%}

%\vspace{5cm}       % Give the correct figure height in cm


%\begin{table}
%\begin{tabular}{lll}
%\hline\noalign{\smallskip}
%\end{tabular}
% Or use
%\vspace*{5cm}  % with the correct table height
%\end{table}
%
% The section below may be edited at your convenience to acknowledge 
% each author's contribution to the manuscript.
% You may remove it if you are a single author.
%
\section{Aknowledgements}
This project was supported by the University of Southampton as host of the Vice-Chancellor Fellowship scheme.

D.R. would like to thank the Ludwig Maximilian University (Munich) for the hospitality during the writing of the thesis that lead to this work.
\section{Author contribution}
Starting from an idea by A. R., D. R. has carried out the computations and produced the data in this work, as well as working towards their interpretation with A. R.\\ 
C. L. has contributed to detailing such interpretations and helped D.R. in writing the paper.\\
D. R., A. R., C. L.  have contributed to the discussion of the results and the revision of the paper.
% BibTeX users please use
 \bibliographystyle{unsrt}
 \bibliography{biblio}
% \begin{thebibliography}{}
%Author, \textit{Book title} (Publisher, place year) page numbers

% etc
%\end{thebibliography}

\end{document}